\documentclass[epj,final]{svjour}

\usepackage{latexsym}
\usepackage{url}
\usepackage{amsfonts}
\usepackage{longtable}

\RequirePackage{graphicx}

\begin{document}

\title{On the Earth's tidal perturbations.II. LARES 2 satellite}

\author{V.G. Gurzadyan\inst{1} \and I. Ciufolini\inst{2,3} \and H.G. Khachatryan\inst{1} \and S. Mirzoyan\inst{1} \and A. Paolozzi\inst{2}
}
\institute{Center for Cosmology and Astrophysics, Alikhanian National Laboratory and Yerevan State University, Yerevan, Armenia  \and Scuola di Ingegneria Aerospaziale, Sapienza Universit\`a di Roma, Rome, Italy \and Innovation Academy for Precision Measurement Science and Technology, Chinese Academy of Sciences, Wuhan, China
}
 
\date{Received: date / Revised version: date}

\abstract{
Laser-ranging satellites have proved their efficiency in high precision testing of General Relativity and constraining modified gravity theories proposed to explain the dark sector and the cosmological tensions. The LARES 2 satellite launched in 2022, is currently providing improved information regarding the frame-dragging effect predicted by General Relativity, as well as the geodesy, thus essentially complementing the data of already existing laser-ranging satellites. The proper knowledge of the Earth's tidal perturbation modes is essential for accurately extracting the sought frame-dragging signal. We present the results of computation of 110 significant modes in Doodson number classification for the parameters of LARES 2 satellite, continuing our previous study on those obtained for the LARES satellite.} 

\PACS{
      {04.80.-y}{Experimental studies of gravity}
}     
\maketitle

\section{Introduction}

The observational data on the dark matter and dark energy, as well as the cosmological tensions, notably, the Hubble tension \cite{R1,Val,R2}, have essentially stimulated the studies on the extensions of the General Relativity (GR) and modified gravity theories. The observational surveys, including of the high-redshift Universe by James Webb Space Telescope, are actively analysed to constrain those models. High precision testing of General Relativity by means of the laser ranging satellites is presenting an alternative and instrumental link to the dark sector models. Namely, the Lense-Thirring effect \cite{CW}, the frame-dragging in the Earth's gravity, is efficiently being studied by laser-ranging satellites \cite{Ciu2007}.  

LARES (LAser RElativity Satellite) satellite, was launched in 2012 and was shown to become the best ever test particle moving on geodesics in the Earth's gravity \cite{Ciu2013}. The results of the analysis of the LARES \cite{Ciu2016}, along with of LAGEOS satellites' data, enabled to show the validity of the GR treatment of the frame-dragging to a few percent accuracy \cite{Ciu2016,Ciu2019}. The Earth's tidal perturbations are among those contributing into the satellite data. Hence the analysis of the tidal modes is an important step in revealing the frame-dragging effect in possibly high precision, as of the difference between the satellite's observed orbital elements and of the modelled ones. The tidal perturbation data along with those of frame-dragging, are then fitted to the residuals of the nodes of LAGEOS (Laser Geodynamic Satellite), LAGEOS 2 and LARES \cite{Ciu2016,Ciu2019}. In \cite{Lares_tides} the tidal modes for the LARES's parameters have been computed. 

LARES 2 satellite \cite{Ciu2017a,Ciu2017b} was launched on 13 July 2022, and currently its data are analysed, both for GR testing  \cite{Ciu2023a,Ciu2023b} and geodesy \cite{Sos,Shao}. Below, we continue the analysis of Earth's tidal perturbations of \cite{Lares_tides}, now regarding the parameters of the LARES 2. The inclination of the orbit of  LARES 2 satellite orbital is chosen to be supplementary to that of LAGEOS, i.e. the sum of their inclinations to be $180^\circ$, in order to compensate the errors in modeling the secular shift of their nodes due to the non-sphericity of the Earth's gravity \cite{Ciu2017a,Ciu2023a}. The released first results on LARES 2 confirmed the GR frame-dragging \cite{Ciu2023b}, and its data are considered a substantial contribution to geodesy, the realization of terrestrial reference frames, the recovery of geocenter motion and more \cite{Sos,Shao}.

\section{Earth's tidal modes}

As in \cite{Lares_tides}, we adopt a model for the gravitational potential of the Earth, the quadrupole as having the main contribution in the departure from sphericity. We follow the notations in \cite{Lares_tides} and concisely present the main formulae.

In standard approach the Earth's gravitational potential is represented via series of  spherical harmonics $Y_{lm}(\phi, \lambda)$ \cite{Br,kaula,carp89,Io,iers2010} 
\begin{eqnarray}
V(r,\phi,\lambda)&=&\frac{GM}{R}\sum_{l=0}^{\infty}\left(\frac{R}{r}\right)^{l+1}\sum_{m=-l}^{l}\frac{w_{lm}Y_{lm}(\phi,\lambda)}{2-\delta_{0m}} \nonumber \\
w^{*}_{lm}&=&w_{l-m}=C_{lm} + iS_{lm}\delta_{0m}
\end{eqnarray}
here $R,M$ are mean equatorial radius and mass of the Earth, respectively, $C_{lm},S_{lm}$ are the Stokes coefficients.

The gravitational perturbations by the Moon and Sun are the dominating ones in inducing Earth's tides. 
The tidal perturbation potential can be written as function of six orbital parameters $\{a,i,e,\omega,\mu,\Omega\}$ of the satellite (Eq.(3.70) in \cite{kaula}), \cite{cata}
\begin{equation}
V_{lm}(a,i,e,\omega,\mu,\Omega)=\frac{GM}{R}\left(\frac{R}{a}\right)^{l+1}\sum_{p=0}^{l}F_{lmp}(i)\sum_{q=-\infty}^{\infty}G_{lpq}(e)S_{lmpq}(\omega,\mu,\Omega,\theta),
\end{equation}
where $a, F_{lmp}(i), G_{lpq}(e)$ are the semi-major axis, inclination and eccentricity functions, respectively, and in the case of our interest, $l=2,\,m=0,\,1,\,2$, have the form  
\begin{eqnarray}
F_{201}(i) &=& \frac{3}{4}sin^{2}i - \frac{1}{2},  \nonumber \\
F_{211}(i) &=& -\frac{3}{2}sin\, i\,\, cos\, i,  \nonumber \\
F_{221}(i) &=& -\frac{3}{2}sin^{2}i,  \nonumber \\
G_{210}(e) &=& \frac{1}{\sqrt{(1-e^2)^3}}.
\end{eqnarray}
The angular perturbation function can be represented in the form \cite{kaula,carp89,iers2010}
\begin{eqnarray}
S_{lmpq}(\omega,\mu,\Omega,\theta) &=& N_{lm}\sum_{i}{u_{lm}k_{lm}(\nu_{i})cos(\nu_{i}\,t+\xi_{lmpq})}, \nonumber \\
\xi_{lmpq} &=& (l-2p)\omega+(l-2p+q)\mu+m(\Omega-\theta)+\eta_{lm}(\nu),\\
N_{lm} &=& \sqrt{\frac{2l+1}{4\pi}\frac{(l-m)!}{(l+m)!}} \nonumber ,
\label{slm}
\end{eqnarray}
where $\omega,\mu,\Omega,\theta$ are the argument of the perigee, mean anomaly, the longitude of the ascending node and the sidereal time, respectively. 

Tidal potential harmonic coefficients $u_{lm}$ are listed in \cite{caed},\cite{cata}. 
Here $k_{lm}(\nu)$ are the Love numbers which weakly depend on the frequency of perturbation mode $\nu$ 
\cite{love},\cite{iers2010}. Love numbers for about 110 modes are listed in 
\cite{iers2010} for $l=0,\,1,\,2$, and can be represented in the form 
\begin{eqnarray}
k_{l} &=& k_{l}^{real} + i\,k_{l}^{imag}  \nonumber \\
\delta k_{lm}(\nu) &=& \delta k_{lm}^{real}(\nu) + i\,\delta k_{lm}^{imag}(\nu) \nonumber \\
k_{lm}(\nu) &=& \left|k_{l} + \delta k_{lm}(\nu)\right|
\end{eqnarray}
The phase $\eta_{lm}(\nu)$ in eq. \ref{slm} can be written as
\begin{equation}
\eta_{lm}(\nu)=\arctan\left(\frac{k_{l}^{imag} + \delta k_{lm}^{imag}(\nu)}{k_{l}^{real} + \delta k_{lm}^{real}(\nu)} \right).
\end{equation}
Doodson's number classification  $\left\{i_1,\ldots,i_6\right\}$ is used to denote the tidal modes \cite{doodson}.
The period of cumulative perturbation by the Moon and the Sun depends on the nodal period of the satellite $\dot{\Omega}$ and periodic behavior of the following six parameters associated to fundamental frequencies and periods: 1) mean Lunar time, 2) Lunar mean longitude, 3) Solar mean longitude, 4) longitude of Lunar perigee, 5) longitude of Lunar ascending node, 6) longitude of Solar perigee. The periods for each of these six parameters are given in \cite{doodson} by means of a dimensionless parameter $\alpha_{i}=\left\{347.80925061,\,13.17639673,\,0.98564734,\,0.11140408,\,0.05295392,\,0.00004707\right\}$ and are defined in the following form 
\begin{equation}
\gamma_{i}=\frac{360}{\alpha_{i}}\,days.
\end{equation}

Then, the frequency of the mode is given as \cite{carp89}

\begin{equation}
\dot{\Gamma}_{j,l,m,p}=\sum_{i=1}^{6}\frac{2{\pi}j_i}{\gamma_i}+m(\dot{\Omega}-\frac{2\pi}{\gamma_1})+(l-2p){\dot{\omega}}+(l-2p+q){\dot{\mu}},
\end{equation}
where $\dot{\Omega},\dot{\omega}$ and $\dot{\mu}$ are the frequencies of the longitude of ascending node, the argument of the perigee and the mean anomaly for the satellite, respectively. 

From the Lagrange equation \cite{kaula} one has the following expression for the ascending node $\Omega$ 
\begin{equation}
\frac{d\Omega}{dt}=\frac{1}{\sqrt{GMa(1-e^2)}}\frac{\partial{V}}{\partial{i}}.
\label{omegadot}
\end{equation}

As seen from this formula, the derivative of the potential $V$ is taken over the inclination parameter $i$ and only the argument of {\it cos} in the potential depends on time, hence, the main perturbation of the longitude of the ascending node is due to the variation of {\it cos}. Then, after integration, in linear approximation one arrives at the following expression for the amplitude of perturbations (cf. Eq. (3.76) in \cite{kaula})

\begin{eqnarray}
\Delta \Omega  &=&\frac{180}{\pi{\sin(i)}}\sum_{l=0}^{\infty }\sum_{m=0}^{l}\sqrt{\frac{GMR^{2l-2}}{a^{2l+3}(1-e^{2})}\frac{2l+1}{4\pi}\frac{(l-m)!}{(l+m)!}}\times   \nonumber \\
&&\times \sum_{p=0}^{l}\sum_{q=-\infty }^{\infty }\frac{dF_{lmp}(i)}{di}G_{lpq}(e)\frac{u_{lm}k_{lm}(\nu )}{\dot{\Gamma}_{j,l,m,p}}.
\end{eqnarray}
 
Thus, our aim is to obtain the modes with long period without mean anomaly of the satellite $\mu$, i.e. at $l=2,p=1,q=0,m=0,1,2$, while other combinations of the parameters are omitted in view of the condition
$$G_{20-2}(e)=G_{222}(e)=0.$$ 

\section{Tidal perturbations of LARES 2}

The parameters of the orbit of the LARES 2 satellite, measured on July 24, 2022, are as follows \cite{Ciu2023b}: 

\begin{eqnarray}
a_{L} &=& 12270210.7550 \,m,  \nonumber \\ 
e_{L} &=& 0.000237,  \nonumber \\
i_{L} &=& 70.15768835^{\circ},  \nonumber \\
P_{L} &=& 225.4\, min,  \nonumber \\
P({\Omega}_{L}) &=& -1050\,days, \nonumber \\
\end{eqnarray}
where $i_L$ is the inclination, i.e. the angle between the orbital plane and the equator, $P_L$ is the orbital period,
$P({\Omega}_{L})$ is the nodal period. The measurement accuracies yield for the semi-major axis,  $\pm 0.1$mm 
i.e. a relative error of $10^{-11}$, for the eccentricity of the orbit, $\pm 1.\, 10^{-6}$, 
for the inclination angle, $\pm (1.\, 10^{-8})^{\circ}$ or less for the mean uncertainty. 
We performed computations for 110 significant tides for these parameters of LARES 2 orbit, the results are given in the Table 1.

\begin{longtable}[B]{lrrrrr}
\multicolumn{6}{c}{{\bf Table 1}. Amplitudes $\Delta\Omega$ and periods of perturbations for the LARES 2 satellite}\\
\multicolumn{6}{c}{generated by Moon and Sun and inducing the tides of the Earth.}\\
\multicolumn{6}{c}{$l=2$,$m=0$,$p=1$,$q=0$}\\
\hline
 & Mode & Love number & Period(days) & $u_{lm}$ & $\Delta\Omega(mas)$ \\
\hline\endfirsthead
 & Mode & Love number & Period(days) & $u_{lm}$ & $\Delta\Omega(mas)$ \\
\hline\endhead
\hline
\multicolumn{6}{c}{\textit{Continued on next page}}
\endfoot
\hline\endlastfoot
             & 055.565 & 0.315416 & 6798.36   &  0.02793 & 1074.57 \\
             & 055.575 & 0.313178 & 3399.18   & -0.00027 &   -5.15708 \\
 $S_a$       & 056.554 & 0.30739  &  365.26   & -0.00492 &   -9.91131 \\
 $S_{sa}$    & 057.555 & 0.305946 &  182.6211 & -0.031   &  -31.0765 \\
             & 057.565 & 0.305896 &  177.844  &  0.00077 &    0.751585 \\
             & 058.554 & 0.305174 &  121.749  & -0.00181 &   -1.20661 \\
 $M_{sm}$    & 063.655 & 0.30292  &   31.8119 & -0.00673 &   -1.16361 \\
             & 065.445 & 0.302709 &   27.6667 &  0.00231 &    0.347112 \\
 $M_m$       & 065.455 & 0.302709 &   27.5546 & -0.03518 &   -5.26489 \\
             & 065.465 & 0.302699 &   27.4433 &  0.00229 &    0.341317 \\
             & 065.655 & 0.302679 &   27.0925 &  0.00188 &    0.276608 \\
 $M_{sf}$    & 073.555 & 0.301818 &   14.7653 & -0.00583 &   -0.466155 \\
             & 075.355 & 0.301728 &   13.7773 & -0.00288 &   -0.214806 \\
 $M_f$       & 075.555 & 0.301718 &   13.6608 & -0.06663 &   -4.927444 \\
             & 075.565 & 0.301718 &   13.6334 & -0.02762 &   -2.03846 \\
             & 075.575 & 0.301718 &   13.6061 & -0.00258 &   -0.190033   \\
 $M_{stm}$   & 083.655 & 0.301257 &    9.55685& -0.00242 &   -0.125009  \\
 $M_{tm}$    & 085.455 & 0.301197 &    9.13293& -0.01276 &   -0.629776 \\
             & 085.465 & 0.301197 &    9.12068& -0.00529 &   -0.26074 \\
 $M_{sqm}$   & 093.555 & 0.300886 &    7.09579& -0.00204 &   -0.0781462 \\
 $M_{qm}$    & 095.355 & 0.300846 &    6.85939& -0.00169 &   -0.0625737 \\
\end{longtable}

\begin{longtable}{lrrrrr}
\multicolumn{6}{c}{{\bf Table 2}. Amplitudes $\Delta\Omega$ and periods of perturbations for the LARES 2 satellite}\\
\multicolumn{6}{c}{generated by Moon and Sun and inducing the tides of the Earth.}\\
\multicolumn{6}{c}{$l=2$,$m=1$,$p=1$,$q=0$}\\
\hline
 & Mode & Love number & Period(days) & $u_{lm}$ & $\Delta\Omega(mas)$ \\
\hline\endfirsthead
 & Mode & Love number & Period(days) & $u_{lm}$ & $\Delta\Omega(mas)$ \\
\hline\endhead
\hline
\multicolumn{6}{c}{\textit{Continued on next page}}
\endfoot
\hline\endlastfoot
$2Q_1$       & 125.755 & 0.298013 &    -6.81487& -0.00664   &     0.238152 \\
$\sigma_1$   & 127.555 & 0.298003 &    -7.04816& -0.00802   &     0.297484 \\
             & 135.645 & 0.297853 &    -9.04214& -0.00947   &     0.450419 \\
$Q_1$        & 135.655 & 0.297843 &    -9.05418& -0.0502    &     2.39075 \\
$\rho_1$     & 137.455 & 0.297813 &    -9.47065& -0.00954   &     0.475188 \\
             & 145.545 & 0.297483 &   -13.4586 & -0.04946   &     3.49712 \\
$O_1$        & 145.555 & 0.297473 &   -13.4853 & -0.26221   &    18.576 \\
$\tau_1$     & 147.555 & 0.297393 &   -14.5605 &  0.00343   &    -0.262298 \\
${N\tau}_1$  & 153.655 & 0.296623 &   -23.4083 &  0.00194   &    -0.237886 \\
             & 155.445 & 0.296373 &   -26.3088 &  0.00137   &    -0.188649 \\
${Lk}_1$     & 155.455 & 0.296363 &   -26.4111 &  0.00741   &    -1.02428 \\
${No}_1$     & 155.655 & 0.296333 &   -26.8499 &  0.02062   &    -2.89737 \\
             & 155.665 & 0.296323 &   -26.9564 &  0.00414   &    -0.58401 \\
$\chi_1$     & 157.455 & 0.295993 &   -30.8765 &  0.00394   &    -0.635913 \\
             & 157.465 & 0.295973 &   -31.0173 &  0.00087   &    -0.141048 \\
$\pi_1$      & 162.556 & 0.289961 &  -109.099  & -0.00714   &     3.98889 \\
             & 163.545 & 0.287131 &  -152.084  &  0.00137   &    -1.05652 \\
$P_1$        & 163.555 & 0.286921 &  -155.565  & -0.12203   &    96.1905 \\
             & 164.554 & 0.28066  &  -270.972  &  0.00103   &    -1.38336 \\
$S_1$        & 164.556 & 0.28066  &  -270.991  &  0.00289   &    -3.88174 \\
             & 165.345 & 0.267821 &  -581.941  &  0.00007   &    -0.19267 \\
             & 165.535 & 0.262000 &  -802.202  &  0.00005   &    -0.185587 \\
             & 165.545 & 0.259851 &  -909.525  & -0.0073    &    30.4688 \\
$K_1$        & 165.555 & 0.257463 & -1050.     &  0.36878   & -1760.61 \\
             & 165.565 & 0.254755 & -1241.79   &  0.05001   &  -279.397 \\
             & 165.575 & 0.251757 & -1519.31   & -0.00108   &     7.29534 \\
             & 166.455 & 1.176597 &   677.473  & -0.0000037 &    -0.05223617 \\
             & 166.544 & 0.650619 &   610.388  &  0.000009  &     0.0637034 \\
$\psi_1$     & 166.554 & 0.526244 &   560.099  &  0.00293   &    15.2515 \\
             & 166.556 & 0.526104 &   560.017  & -0.000042  &     -0.219104 \\
             & 166.564 & 0.466726 &   517.467  &  0.00005   &      0.213259 \\
             & 167.355 & 0.335867 &   256.113  &  0.00018   &      0.273441 \\
             & 167.365 & 0.333836 &   246.815  &  0.00006   &      0.0873069 \\
$\phi_1$     & 167.555 & 0.328564 &   221.071  &  0.00525   &      6.73447 \\
             & 167.565 & 0.327234 &   214.108  & -0.0002    &     -0.247465 \\
             & 168.554 & 0.314689 &   137.718  &  0.00031   &      0.237261 \\
$\theta_1$   & 173.655 & 0.302004 &    32.8059 &  0.00395   &      0.691122 \\
             & 173.665 & 0.301994 &    32.6483 &  0.00078   &      0.135815 \\
             & 175.445 & 0.301554 &    28.4154 & -0.0006    &     -0.0907953 \\
$J_1$        & 175.455 & 0.301544 &    28.2971 &  0.02062   &      3.10724 \\
             & 175.465 & 0.301534 &    28.1798 &  0.00409   &      0.61375 \\
${So}_1$     & 183.555 & 0.300244 &    14.9759 &  0.00342   &      0.271572 \\
             & 185.355 & 0.300154 &    13.9605 &  0.00169   &      0.125061 \\
${Oo}_1$     & 185.555 & 0.300144 &    13.8409 &  0.01129   &      0.828284 \\
             & 185.565 & 0.300144 &    13.8127 &  0.00723   &      0.529347 \\
             & 185.575 & 0.300144 &    13.7847 &  0.00151   &      0.110331 \\
$\nu_1$      & 195.455 & 0.299714 &     9.21307&  0.00216   &      0.105331 \\
             & 195.465 & 0.299714 &     9.2006 &  0.00138   &      0.0672039 \\
\end{longtable}

\newpage

\begin{longtable}{lrrrrr}
\multicolumn{6}{c}{{\bf Table 3}. Amplitudes $\Delta\Omega$ and periods of perturbations for the LARES 2 satellite}\\
\multicolumn{6}{c}{generated by Moon and Sun and inducing the tides of the Earth.}\\
\multicolumn{6}{c}{$l=2$,$m=2$,$p=1$,$q=0$}\\
\hline
 & Mode & Love number & Period(days) & $u_{lm}$ & $\Delta\Omega(mas)$ \\
\hline\endfirsthead
 & Mode & Love number & Period(days) & $u_{lm}$ & $\Delta\Omega(mas)$ \\
\hline\endhead
\hline
\multicolumn{6}{c}{\textit{Continued on next page}}
\endfoot
\hline\endlastfoot
       & 225.855 & 0.301083 &   -5.43532&  0.0018   &  -0.0215766 \\
       & 227.655 & 0.301083 &   -5.58269&  0.00467  &  -0.0574971 \\
       & 235.755 & 0.301083 &   -6.77093&  0.01601  &  -0.23907 \\
       & 237.555 & 0.301083 &   -7.00117&  0.01932  &  -0.298306 \\
       & 245.555 & 0.301083 &   -8.9519 & -0.00389  &   0.076798 \\
       & 245.645 & 0.301083 &   -8.96493& -0.00451  &   0.089168 \\
 $N_2$ & 245.655 & 0.301083 &   -8.97677&  0.12099  &  -2.39527 \\
       & 247.455 & 0.301063 &   -9.386  &  0.02298  &  -0.475649 \\
       & 253.755 & 0.301063 &  -12.5056 & -0.0019   &   0.0523982 \\
       & 254.556 & 0.301063 &  -12.8461 & -0.00218  &   0.0617566 \\
       & 255.545 & 0.301063 &  -13.2883 & -0.02358  &   0.690987 \\
 $M_2$ & 255.555 & 0.301063 &  -13.3143 &  0.63192  & -18.554 \\
       & 256.554 & 0.301063 &  -13.818  &  0.00192  &  -0.0585065 \\
       & 263.655 & 0.301063 &  -22.8978 & -0.00466  &   0.235308 \\
       & 265.455 & 0.301063 &  -25.763  & -0.01786  &   1.01469 \\
       & 265.555 & 0.301063 &  -25.9701 &  0.00359  &  -0.2056 \\
       & 265.655 & 0.301063 &  -26.1805 &  0.00447  &  -0.258072 \\
       & 265.665 & 0.301063 &  -26.2817 &  0.00197  &  -0.114176 \\
       & 271.557 & 0.301063 &  -77.7839 &  0.0007   &  -0.120073 \\
 $T_2$ & 272.556 & 0.301063 &  -98.8303 &  0.0172   &  -3.74865 \\
 $S_2$ & 273.555 & 0.301063 & -135.491  &  0.294    & -87.8442 \\
       & 273.755 & 0.301063 & -147.892  &  0.00004  &  -0.0130455  \\
       & 274.554 & 0.301063 & -215.387  & -0.00246  &   1.16845 \\
       & 274.556 & 0.301063 & -215.399  &  0.00062  &  -0.294505 \\
       & 274.566 & 0.301063 & -222.447  & -0.00004  &   0.019622 \\
       & 273.655 & 0.301063 & -141.42   &  0.00004  &  -0.0124746 \\
       & 275.455 & 0.301063 & -451.627  &  0.00019  &  -0.18923 \\
       & 275.465 & 0.301063 & -483.764  &  0.00004  &  -0.0426727 \\
       & 275.545 & 0.301063 & -487.364  &  0.00103  &  -1.107  \\
 $K_2$ & 275.555 & 0.301063 & -525.000  &  0.07996  & -92.5739  \\
       & 275.565 & 0.301063 & -568.936  &  0.02383  & -29.8981  \\
       & 277.555 & 0.301063 &  280.029  &  0.00063  &   0.389045 \\
       & 275.575 & 0.301063 & -620.897  &  0.00259  &  -3.5463 \\
       & 282.656 & 0.301063 &   37.3243 &  0.00004  &   0.00329236 \\
       & 283.445 & 0.301063 &   34.7657 &  0.00006  &   0.000001 \\
       & 283.455 & 0.301063 &   34.5867 &  0.00004  &   0.0031 \\
       & 283.655 & 0.301063 &   33.8631 &  0.00085  &   0.0634764 \\
       & 283.665 & 0.301063 &   33.696  &  0.00037  &   0.0274939 \\
       & 283.675 & 0.301063 &   33.5299 &  0.00004  &   0.00295766  \\
       & 285.455 & 0.301063 &   29.0809 &  0.00447  &   0.286663 \\
       & 285.465 & 0.301063 &   28.957  &  0.00195  &   0.124521 \\ 
\end{longtable}

Figures 1, 2 represent the frequency counts for the amplitude, period for the 110 tidal modes and Figure 3 exhibits the amplitude vs period.

\begin{figure}[!htbp]
  \centering
  \includegraphics[width=100mm]{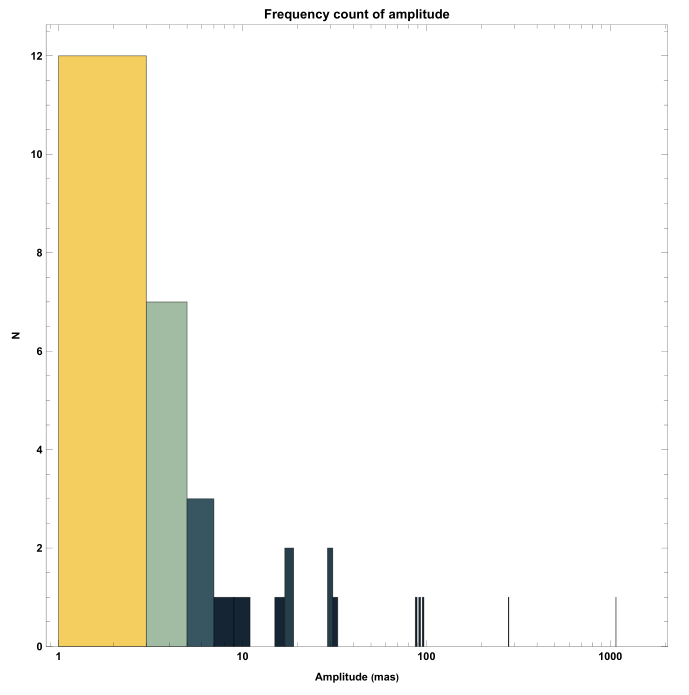}
  \label{fig:1}
  \caption{The frequency count for the amplitudes (in milliarcseconds: {\it mas}) of 110 significant tidal modes of the LARES-2 satellite.}
\end{figure}

\begin{figure}[!htbp]
  \centering
  \includegraphics[width=100mm]{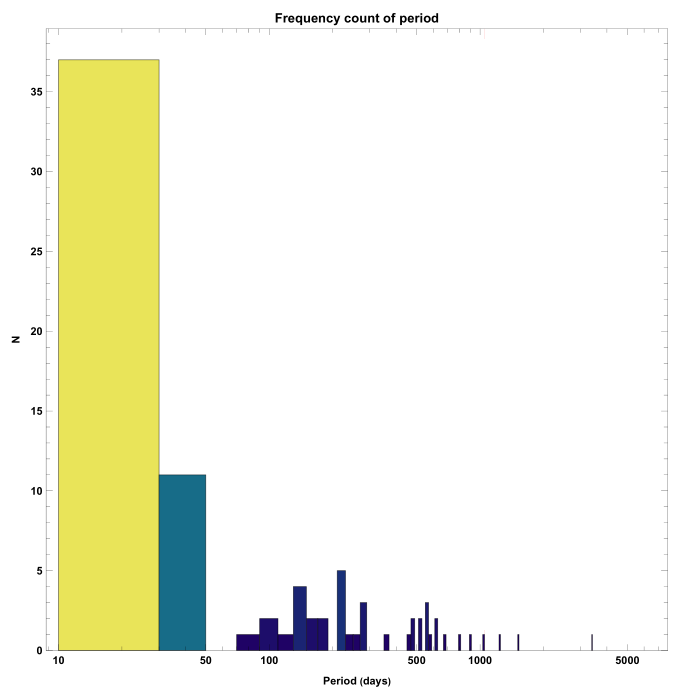}
  \label{fig:2}
  \caption{The frequency count for the periods (in days) of 110 tidal modes for the LARES-2 satellite.}
\end{figure}

\begin{figure}[!htbp]
  \centering
  \includegraphics[width=100mm]{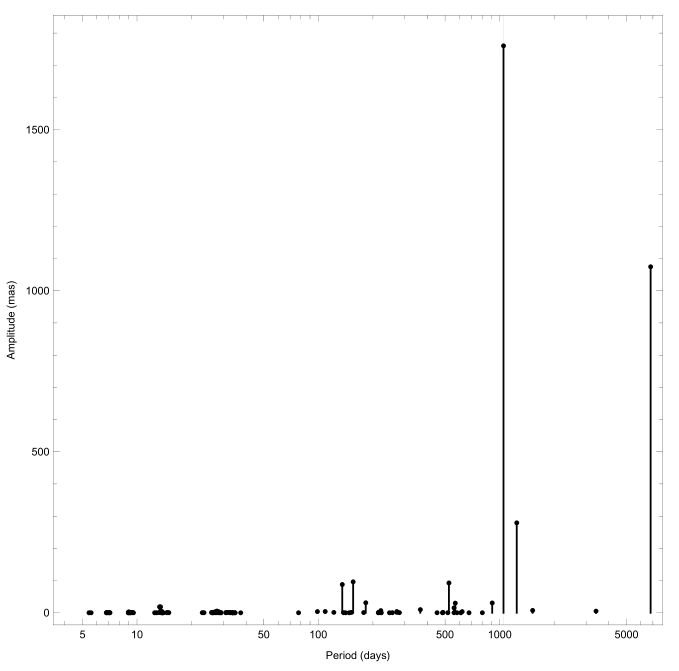}
  \label{fig:3}
  \caption{The amplitude vs period for the modes of Figures 1 and 2.}
\end{figure}

\section{Conclusions}

We computed the 110 significant Earth's tidal modes for the LARES 2 satellite using the perturbative methods of celestial mechanics and the data on the satellite's orbit. The modes are presented according the Doodson number classification, and their relative relevance is made explicit.

The laser ranging data on LARES 2 satellite in combination with the data of two LAGEOS and LARES satellites, taking into account the parameters of the Earth's tidal modes, and at data storage over a sufficiently long time period to guarantee adequate statistic, lead to a high precision testing of frame-dragging and, hence, of General Relativity \cite{Ciu2023a}. Such studies are complementing the ongoing astrophysical surveys in constraining the modified gravity models proposed to explain the dark sector, including e.g. the role of the cosmological constant in the local Universe \cite{GS1,GS2}. Note, that recent efforts to test gravity and the cosmological constant include quantum technologies \cite{BEC}.  

\section{Acknowledgments}
We thank the referee for valuable comments. We acknowledge the Italian Space Agency for the support of the LARES 2 space mission under agreement No.2024-3-HH.0.  We also acknowledge the Chinese Academy of Sciences President’s International Fellowship Initiative, grant
No. 2025PVA0049. We are also grateful to the European Space Agency, International Ranging Service,  AVIO and Spacelab S.p.A. 

\section{Data Availability Statement} 
The data supporting the findings of this study are available in the paper.

\end{document}